\documentclass[letterpaper, 12pt, onecolumn,final,conference]{ieeeconf}

    \IEEEoverridecommandlockouts            
    
    \usepackage{cite}      
    \usepackage{amsfonts}
    \usepackage{amssymb}
    \usepackage{amstext}
    \usepackage{amsmath}
    \usepackage{graphicx}
    \usepackage{enumerate}
    \usepackage{tikz}
    \usepackage{pgfplots}
    \usepackage{bm}
    \usepackage{bbold}
    \usepackage{etoolbox}
    \usepackage{cancel}
    \usepackage{changes}

    \patchcmd{\subequations}{\def\theequation{\theparentequation\alph{equation}}}%
    {\def\theequation{\theparentequation.\arabic{equation}}}{}{}

    \graphicspath{{figs/}}

    \newtheorem{theorem}{Theorem}[section]

    \newtheorem{remark}[theorem]{Remark}

    \definechangesauthor[color=red]{FS}
    \definechangesauthor[color=blue]{GN}
    \definechangesauthor[color=magenta]{AC}

    \def \final_ver{0}
    
    \newcommand{\RM}[2]{
        \ifnum \final_ver = 1
        \ignorespaces{#2}
        \else
        \deleted[id=#1]{#2}
        \fi
    }
    
    \newcommand{\AD}[2]{
        \ifnum \final_ver = 1
        {#2}
        \else
        \added[id=#1]{#2}
        \fi
    }

    \newcommand{\idxl}{\ell}
    \newcommand{\idxm}{m}
    
    \newcommand{\idxi}{i}
    \newcommand{\idxj}{j}
    
    \newcommand{\idxL}{L}
    \newcommand{\idxI}{N}
    \newcommand{\nTesting}{M}

    \newcommand{\randomO}{X}
    \newcommand{\sampleO}{x}
    \newcommand{\sampleOagent}{\bm{x}}
    \newcommand{\sampleOall}{\bm{\mathrm{x}}}
    \newcommand{\meanO}{\overline{x}}
    \newcommand{\meanOall}{\overline{\bm{\mathrm{x}}}}
    \newcommand{\spaceO}{\randomO}

    \newcommand{\sampleP}{\theta}
    \newcommand{\samplePtesting}{\theta^R}
    \newcommand{\samplePall}{\bm{\theta}}
    \newcommand{\samplePtestingAll}{\bm{\theta}^R}

    \newcommand{\prob}{p}

    \newcommand{\hyper}{\bm{\gamma}}
    \newcommand{\hyperS}{\Gamma}
    
    \newcommand{\varianceO}{\sigma^2}
    
    \newcommand{\meanP}{\mu}
    \newcommand{\meanPall}{\bm{\mu}}
    \newcommand{\meanPtesting}{\mu^R}
    \newcommand{\meanPtestingAll}{\bm{\mu}^R}
    \newcommand{\meanPallHyper}{\meanPall_{\hyper}}
    \newcommand{\meanPtestingAllHyper}{\meanPtestingAll_{\hyper}}
    \newcommand{\covarianceP}{K}
    
    \newcommand{\spatial}{s}
    \newcommand{\spatialBold}{\bm{s}}
    
    \newcommand{\spatialS}{P}
    \newcommand{\testing}{s^R}
    
    \newcommand{\z}{\bm{z}}
    \newcommand{\zScalar}{z}

    \newcommand{\diag}{D}
    
    \newcommand{\graph}{\mathcal{G}}
    \newcommand{\edges}{\mathcal{E}}
    \newcommand{\neigh}{\mathcal{N}}
    
    \newcommand{\comm}{{\texttt{cmm}}}
    \newcommand{\inter}{{\texttt{int}}}
    
    \newcommand{\interp}{\mathcal{I}}

    \def \jphml/{JPH-ML}

    \DeclareMathOperator*{\argmax}{argmax}
    \DeclareMathOperator*{\subj}{subj. to}

    \newcommand\oprocendsymbol{\hbox{$\square$}}
    \newcommand\oprocend{\relax\ifmmode\else\unskip\hfill\fi\oprocendsymbol}
    \def\eqoprocend{\tag*{$\square$}}
    
    \renewcommand{\theenumi}{(\roman{enumi})}

    \title{An Empirical Bayes Approach for Distributed Estimation of Spatial Fields}

    \author{Francesco Sasso, Angelo Coluccia, {\it Senior Member, IEEE}\\ and Giuseppe
        Notarstefano, {\it Member, IEEE} 
        \thanks{Francesco Sasso, Angelo Coluccia and Giuseppe Notarstefano are with the Department of
            Engineering, Universit\`a del Salento, via Monteroni, 73100, Lecce, Italy, \{name.lastname\}@unisalento.it. 
        }  
        \thanks{This result is part of a project that has received funding from the European Research Council
            (ERC) under the European Unions Horizon 2020 research and innovation
            programme (grant agreement No 638992 - OPT4SMART).
        }
    }

\begin{document}
        \bstctlcite{IEEEexample:BSTcontrol}
        \maketitle

        \begin{abstract}
            In this paper we consider a network of spatially distributed sensors which
            collect measurement samples of a spatial field, and aim at estimating in a
            distributed way (without any central coordinator) the entire field by suitably
            fusing all network data. We propose a general probabilistic model that can
            handle both partial knowledge of the physics generating the spatial field as
            well as a purely data-driven inference. Specifically, we adopt an Empirical
            Bayes approach in which the spatial field is modeled as a Gaussian Process,
            whose mean function is described by means of parametrized equations.  We
            characterize the Empirical Bayes estimator when nodes are heterogeneous, i.e.,
            perform a different number of measurements. Moreover, by exploiting the
            sparsity of both the covariance and the (parametrized) mean function of the
            Gaussian Process, we are able to design a distributed spatial field estimator.
            We corroborate the theoretical results with two numerical simulations: a
            stationary temperature field estimation in which the field is described by a
            partial differential (heat) equation, and a data driven inference in which the
            mean is parametrized by a cubic spline.
        \end{abstract}

        \section{Introduction}
        The presence of ubiquitous portable devices makes available a massive amount of spatially
        distributed measurements of several quantities, which can be used to estimate spatial fields of interest. While single measurements at each
        node can be inaccurate, fusion of data from multiple nodes gives the possibility to
        have a much more reliable estimate of the field. In order to avoid collecting
        and processing all data in a single computing unit, distributed estimation
        methods play an important role. Indeed, through this new computation paradigm,
        devices can improve their local measurement of the field and predict nearby
        values.
        
        There are two main batches of literature
        related to the set-up investigated in this paper, namely Gaussian Process
        Regression and Kriging.  
        Gaussian Process Regression has received significant attention in Artificial
        Intelligence and Machine Learning \cite{rasmussen2006gaussian}. 
        In a distributed context, efficient Gaussian Process Regression is
        discussed in \cite{carron2016machine,todescato2017distributed}, while
        \cite{xu2011bayesian} presents a Bayesian approach for (distributed)
        spatio-temporal regression.
        Kriging interpolation techniques have been widely investigated in geostatistics
        \cite{cressie2015statistics}, \cite{stein2012interpolation}.
        In this context, one of the first references on distributed estimation of a spatial field is
        \cite{cortes2009distributed}, which formulates the estimation of a
        spatio-temporal field via Bayesian Universal Kriging. Reference~\cite{martinez2010distributed} presents adaptive
        interpolation schemes for time-varying field estimation. In \cite{varagnolo2012distributed} distributed estimators are proposed
        for regression problems in sensor networks.
        In \cite{kozma2012distributed} a receding horizon approach is  considered for
        estimation of spatially distributed systems modeled through an advection-diffusion
        Partial Differential Equation (PDE). A similar PDE is considered in \cite{georges2017optimal} where the
        optimal sensor location problem is investigated for pollution monitoring. 
        In \cite{elwin2016environmental} an environmental estimation method is developed.
        Finally, distributed estimation of a spatial field has application in
        cooperative control as highlighted in \cite{graham2010spatial,le2009trajectory,bell2009distributed}.

        The contribution of the paper is twofold. First, we propose a probabilistic
        model for a network estimation set-up in which sensors take
        (possibly multiple) measurements of a spatial field and aim at cooperatively
        estimating the overall field by optimally fusing all data.
        Specifically, we propose an Empirical Bayesian framework in which the spatial
        field is modeled as a Gaussian Process having known covariance and, differently
        from existing works, non-zero mean depending on some hyperparameters that are
        estimated from the all measurements.
        In particular, the mean of the process is a parametric function that might
        encode some (deterministic) physical law modeling the expected behavior of the
        spatial field.
        This probabilistic model extends the one introduced in
        \cite{coluccia2013distributed,coluccia2016bayesian} where the prior distribution
        did not depend on the spatial displacement of the nodes.
        Second, we design a distributed estimator of the spatial field relying on a
        Maximum Likelihood (ML) estimator of the hyperparameters and a local Maximum A
        Posteriori (MAP) estimator. We show how the ML estimator can be obtained by
        solving a structured optimization problem amenable to distributed
        computation. In particular, the optimization problem has a partitioned structure
        which calls for more efficient distributed optimization algorithms.

        The paper is organized as follows. In Section~\ref{sec:setup} we present the
        spatial field estimation set-up.  In Section~\ref{sec:empirical_bayes} we
        introduce the proposed probabilistic model and give two example scenarios.  In
        Section~\ref{sec:distributed_estimator} we derive the distributed estimator and,
        finally, in Section~\ref{sec:simulations} we provide numerical simulations to
        show the performance of the distributed estimator applied to the two examples in
        Section~\ref{sec:empirical_bayes}.  .

        \section{Spatial Field Estimation Set-up}
        \label{sec:setup}
\begin{figure}
            \centering
            \includegraphics[scale=0.6]{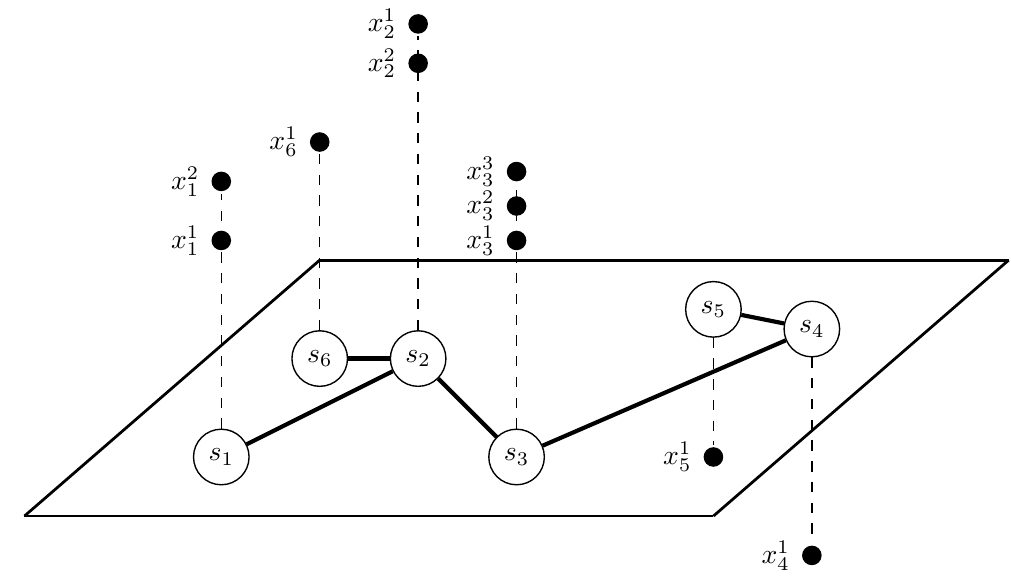}
            \caption{Example of graph $\graph^\comm$ and its related observations $\sampleOall$.}
            \label{fig:network-measurements}
        \end{figure}

        We consider a network of $\idxI$ spatially distributed sensors with local
        computation and communication capabilities. Nodes communicate according to a
        \emph{communication graph} $\graph^\comm = (\{1,\dots,\idxI\}, \edges^\comm)$.
        Each sensor has the possibility to perform multiple measurements of an external
        quantity at the point where it is located as represented in
        Fig.~\ref{fig:network-measurements}. The goal for the network is to suitably
        fuse the spatially distributed measurements in order to estimate the spatial
        field.
        Formally, each sensor $\idxi = 1,\dots,\idxI$:
        \begin{itemize}
            \item is at a \emph{measurement} (or \emph{training}) point $\spatial_\idxi\in\spatialS\subset\mathbb{R}^d$;
            \item makes $\idxL_\idxi$ \emph{observations} $\sampleO^\idxl_\idxi\in\spaceO\subset\mathbb{R}$, $\idxl=1,\dots,\idxL_\idxi$.
        \end{itemize}
        Each observation $\sampleO^\idxl_\idxi$ is obtained as:
        \begin{align}\label{eq:observations}
        \sampleO^\idxl_\idxi = \sampleP(\spatial_\idxi) +
        \epsilon_\idxi^\idxl, %
        \end{align}
        where all $\epsilon_\idxi^\idxl$ are i.i.d. (noise) random variables with
        Gaussian distribution $\mathcal{N}(0,\varianceO)$, while
        $\sampleP:\spatialS\mapsto\spaceO$ is an unknown vector valued function.
        
        The unknown function $\sampleP(\cdot)$ represents a \emph{spatial field} that
        sensors sample through noisy observations at the measurement points
        $\spatial_\idxi$, with $\epsilon_\idxi^\idxl$ being the measurement noise.
        
        For notational purpose, we introduce the following shorthands:
        \begin{equation*}
        \begin{split}
        \sampleOagent_\idxi &= (\sampleO_\idxi^1,\dots,\sampleO_\idxi^{\idxL_\idxi})^\top,\quad\\
        \sampleP_\idxi &= \sampleP(\spatial_\idxi),
        \end{split}
        \quad
        \begin{split}
        \sampleOall &=
        (\sampleOagent_1^\top,\dots,\sampleOagent_\idxI^\top)^\top,\\ 
        \samplePall &= (\sampleP_1,\dots,\sampleP_\idxI)^\top.
        \end{split}
        \end{equation*}
        From the observation model \eqref{eq:observations} and the assumption that
        noises are i.i.d. Gaussian variables, it follows that
        \begin{subequations}
            \begin{align}
            \prob(\sampleOall|\samplePall) &=
            \prod_{\idxi=1}^\idxI\prob(\sampleOagent_\idxi|\sampleP_\idxi)
            =
            \prod_{\idxi=1}^\idxI\prod_{\idxl=1}^{\idxL_\idxi}\prob(\sampleO_\idxi^\idxl|\sampleP_\idxi),\label{eq:joint-assumption1}\\
            p(\sampleO_\idxi^\idxl | \sampleP_\idxi) &\sim \mathcal{N}(\sampleP_\idxi,\varianceO).\label{eq:joint-assumption2}
            \end{align}
        \end{subequations}

        Given the observations $\sampleOall$ and the graph $\graph^\comm$, our goal is
        to design a distributed algorithm allowing nodes to estimate the value of
        the spatial field $\sampleP(\cdot)$ in a set of \emph{regression} (or
        \emph{testing}) points $\testing_1,\dots,\testing_{\nTesting}\in\spatialS$.
        
        It is worth noticing that the regression points can either coincide or not with
        the measurement points. In the first case, a node $\idxi$ in the network is
        trying to improve its local estimate of $\sampleP(\spatial_\idxi)$, while in the
        second one it wants to interpolate the field in a neighboring point where
        measurements are missing.
        In the following we will show how a single node can take advantage of all other
        measurements in the network by exchanging information only with neighboring
        nodes.
        \section{Empirical Bayes Framework}
        \label{sec:empirical_bayes}
        In this section we introduce a probabilistic model allowing nodes, especially
        the ones with few measurements, to have a better estimate of the spatial field
        at their measurement points or in neighboring areas.

        \subsection{Probabilistic Model}
        We adopt a Bayesian model in which the spatial field $\sampleP(\cdot)$ is a
        \emph{Gaussian Process} with mean function $\meanP(\cdot)$ and covariance function
        $\covarianceP(\cdot,\cdot)$, denoted by
        \begin{align}
        \sampleP(\cdot) &\sim
        \mathcal{GP}(\meanP(\cdot),\covarianceP(\cdot,\cdot)).
        \label{eq:joint-assumption3}
        \end{align}

        For the covariance function $\covarianceP(\cdot,\cdot)$, known \emph{kernel}
        functions can be used as customary.
        As for the mean function $\meanP(\cdot)$, different from existing works where
        it is assumed to be null, we consider a more general parametric model which is
        meant to capture some prior knowledge. Since it is not realistic to assume the
        model (hyper)parameters to be known by nodes in the network, we propose an
        approach, based on the Empirical Bayes paradigm, in which hyperparameters will
        be estimated from the data through Maximum Likelihood (ML).
        As we will show through two example scenarios, this
        framework can model both a set-up in which prior knowledge encodes some
        (deterministic) physical law (but with unknown values of the physical
        parameters), and a data-driven inference scenario in which an interpolating
        function is used.
        
        Formally, we introduce an interaction graph,
        $\graph^\inter=(\{1,\dots,\idxI\},\edges^\inter)$, induced by the Gaussian
        Process. We denote $\neigh_\idxi^\inter = \{\idxj :
        (\idxi,\idxj)\in\edges^\inter\}$ the set of neighbors of node $i$ in the
        interaction graph. Consistently, we let $\spatialBold_{\neigh_\idxi^\inter} =
        \big[\spatial_\idxj\big]_{\idxj\in\neigh_\idxi^\inter}$ be the vector of
        measurement points which are neighbors of point $\spatial_\idxi$, and
        $\meanPall_{\neigh_\idxi^\inter} =
        \big[\meanP(\spatial_\idxj)\big]_{\idxj\in\neigh_\idxi^\inter}$ the vector of
        values of $\meanP(\cdot)$ at those points.
        
        We, thus, assume that
        \begin{align}\label{eq:sparsity}
        \covarianceP(\spatial_\idxi,\spatial_\idxj) \neq 0 \iff (\idxi,\idxj)\in\edges^\inter,
        \end{align}
        and that $\meanP(\cdot)$ is the (unique) solution of the following parametrized system of equations, termed \emph{spatial dynamics}:
        \begin{align}
        F_\idxi(\spatialBold_{\neigh_\idxi^\inter},\meanPall_{\neigh_\idxi^\inter};\hyper) = 0,\qquad\idxi=1,\dots,\idxI,
        \label{eq:joint-assumption4}
        \end{align}
        where the functions $F_\idxi(\cdot,\cdot;\hyper)$ are known, while
        $\hyper\in\hyperS$ is the unknown \emph{hyperparameter} vector.
        
        We point out that $\samplePall$ is the vector of parameters of $\sampleOall$,
        and $\hyper$ is the parameter of $\samplePall$, and thus the name
        hyperparameter.

        \subsection{Examples of application scenarios}\label{subsec:examples}
        Here, we present two scenarios that highlight the flexibility of the proposed
        probabilistic framework. In the first one, we consider a set-up in which
        physical knowledge of the spatial field is available through a partial
        differential equation that can be discretized at the measurement points. In the
        second one, we show how interpolating curves can be used when there is no
        physical knowledge of the phenomenon.
        
        \subsubsection{Temperature Dynamics}
        We consider a scenario in which sensors want to estimate the temperature in a
        given environment. In this case, the mean function $\meanP(\cdot)$ of the
        spatial field $\sampleP(\cdot)$ is expected to obey the following Heat Equation
        parametrized by $\hyper$:
        \begin{align*}
        \sum_{\idxl=1}^d\!\partial^2_{\idxl\idxl}\meanP(\spatial, t)
        - \partial_t\meanP(\spatial, t) = w(\spatial, t;\hyper),
        \end{align*}
        where $(\spatial,t)\!\in\!\spatialS\!\times[0,T)$,
        $\partial_{\idxl}$ is the partial derivative with respect to the $\ell$-th
        component of $\spatial$, and $\partial_t$ is the partial derivative with respect
        to time $t$. The function $w(\cdot;\hyper)$ is also known as the heat source.
        
        If one considers a thermostatic (equilibrium) condition, the Heat Equation turns
        out to be the so called Poisson Equation:
        \begin{align*}
        \sum_{\idxl=1}^d\partial^2_{\idxl\idxl}\meanP(\spatial) =
        w(\spatial;\hyper),\quad\spatial\in\spatialS\subset\mathbb{R}^d.
        \end{align*}
        Now, in order to guarantee the uniqueness of $\meanP(\cdot)$, we specify the
        boundary condition. 
        For the sake of clarity, we consider the 1-dimensional case in which we want to
        monitor the temperature in a bar
        $\spatialS = [\spatial_1,\spatial_\idxI]$. As a consequence,
        $\meanP(\cdot)$ is uniquely determined by the following Cauchy Problem
        \begin{align*}
        \begin{cases} 
        \meanP''(\spatial) = w(\spatial;\hyper), & \quad\spatial\in(\spatial_1,\spatial_\idxI), \\ 
        \hspace{2pt}\meanP(\spatial_1) = w_1, \\
        \meanP(\spatial_\idxI) = w_\idxI. 
        \end{cases}
        \end{align*}
        
        Then, the Poisson Equation can be discretized at the measurement points
        $\spatial_1,\dots,\spatial_\idxI$, by using a suitable approximation rule as,
        e.g.,
        \begin{align*}
        \meanP(\spatial_\idxi)'' \approx \frac{\meanP(\spatial_{\idxi+1}) -
            2\meanP(\spatial_\idxi) + \meanP(\spatial_{\idxi-1})}{\epsilon^2},\quad \idxi
        = 2,\dots,\idxI-1, 
        \end{align*}
        where we have assumed, for simplicity, that the measurement points are uniformly spaced, i.e., $\epsilon \!=\! \spatial_{\idxi+1} - \spatial_{\idxi}$ for
        all $\idxi \!=\! 1,\dots,\idxI-1$. 
        By plugging in the boundary conditions, the discretized system turns out to be
        \begin{align*} 
        \begin{bmatrix} 
        1 & & & & \\ 
        1 & -2 & 1 & & \\ 
        & \ddots & \ddots & \ddots & \\
        & & 1 & -2 & 1 \\
        & & & & 1
        \end{bmatrix}
        \begin{bmatrix}
        \meanP(\spatial_1)\\
        \meanP(\spatial_2)\\
        \vdots\\
        \meanP(\spatial_{\idxI-1})\\
        \meanP(\spatial_\idxI)
        \end{bmatrix} =
        \begin{bmatrix}
        w_1\\
        \epsilon^2 w(\spatial_2;\hyper)\\
        \vdots\\ 
        \epsilon^2 w(\spatial_{\idxI-1};\hyper)\\
        w_\idxI
        \end{bmatrix},
        \end{align*}
        which has the sparsity of \eqref{eq:joint-assumption4} once we set
        \begin{align*}
        \neigh_\idxi^\inter &=
        \begin{cases}
        \{\idxi\}, & \text{for } \idxi\in\{1,\idxI\},\\
        \{\idxi-1,\idxi,\idxi+1\} & \text{otherwise}, 
        \end{cases}
        \end{align*}
        and
        \begin{align*}
        F_\idxi(\spatialBold&_{\neigh_\idxi^\inter},\meanPall_{\neigh_\idxi^\inter};\hyper) =\\ 
        &
        \begin{cases}
        \meanP(\spatial_\idxi) - u_\idxi, & \text{for } \idxi\in\{1,\idxI\},\\
        \frac{\meanP(\spatial_{\idxi+1}) - 2\meanP(\spatial_\idxi) +
            \meanP(\spatial_{\idxi-1})}{\epsilon^2} - u(\spatial_\idxi;\hyper) & \text{otherwise.}
        \end{cases}
        \end{align*}
        
\begin{figure}
            \centering
            \includegraphics[scale=.7]{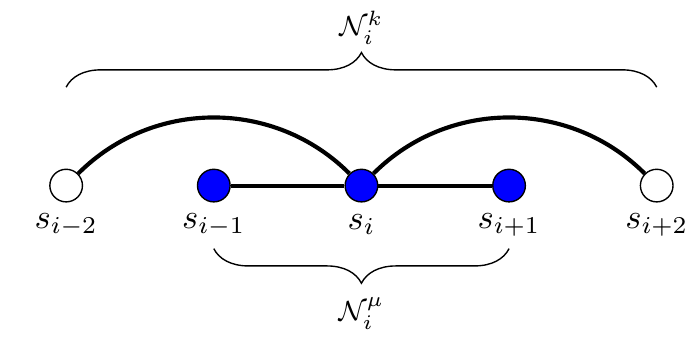}
            \caption{Neighborhoods of agent $\idxi$ induced by the covariance
                function $\covarianceP(\cdot,\cdot)$ and by the mean function
                $\meanP(\cdot)$, denoted by $\mathcal{N}_i^k$ and $\mathcal{N}_i^\mu$
                respectively.}\label{fig-network-poisson}
        \end{figure}

        We point out that, as depicted in Fig. \ref{fig-network-poisson}, the
        discretization process may induce a set of neighbors, call it
        $\neigh_\idxi^\meanP$, while the covariance function may correlate with a
        broader set of agents, call it
        $\neigh_\idxi^{\covarianceP}\supset \neigh_\idxi^\meanP$. Clearly, we can set
        $\neigh_\idxi^\inter = \neigh_\idxi^\covarianceP$.
        
        \subsubsection{Interpolating Dynamics}
        Here we consider a data-driven inference scenario in which the shape of the mean
        $\meanP(\cdot)$ is described by
        \begin{align*}
        \meanP(\spatial) = \interp(\spatial;\hyper),\quad\spatial\in\spatialS,
        \end{align*}
        where $\interp(\spatial;\hyper)$ is an interpolation function (e.g., a cubic
        spline) of the points $\{(\spatial_\idxi,\gamma_\idxi)\}_{\idxi}$, parametrized
        by $\hyper = (\gamma_1,\dots,\gamma_\idxI)^\top\in\mathbb{R}^\idxI$.
        This gives the possibility to fit the observations even if there is no
        information about the physics of the problem.

        \section{Distributed Empirical Bayes Estimator}
        \label{sec:distributed_estimator}
        In this section we develop our distributed estimation approach following the
        Empirical Bayes framework introduced in the previous section. Specifically, we
        derive the Maximum Likelihood (ML) estimator of the hyperparameter and the
        Maximum A Posteriori (MAP) estimator of the spatial field. Then, we show how to
        recast the ML problem into an equivalent optimization problem with a partitioned
        structure that is amenable to distributed computation. Finally, we provide a
        local procedure allowing each node to compute the MAP estimator of the field at
        its location or to perform a regression at nearby points.

        We start by introducing some useful notation:
        \begin{equation*}
        \begin{split}
        \meanP_\idxi &= \meanP(\spatial_\idxi),\\
        \meanPtesting_\idxm &= \meanP(\testing_\idxm),
        \end{split}
        \qquad
        \begin{split}
        \meanPall &= (\meanP_1,\dots,\meanP_\idxI)^\top,\\
        \meanPtestingAll &= (\meanPtesting_1,\dots,\meanPtesting_\nTesting)^\top.
        \end{split}
        \end{equation*}
        When needed, we highlight the dependence of $\meanPall$ and $\meanPtestingAll$
        on $\hyper$ (from \eqref{eq:joint-assumption4}) by  writing $\meanPallHyper$
        and $\meanPtestingAllHyper$ respectively.  
        Moreover, we define the matrices
        $K_{\spatial\spatial}\in\mathbb{R}^{\idxI\!\times\!\idxI}$ with entries
        $[K_{\spatial\spatial}]_{\idxi\idxj} \!= \covarianceP(\spatial_\idxi,\spatial_\idxj)$, %
        $K_{\testing\spatial}\in\mathbb{R}^{\nTesting\!\times\!\idxI}$ with entries
        $[K_{\testing\spatial}]_{\idxm\idxi} \!= \covarianceP(\testing_\idxm,\spatial_\idxi)$, and
        consistently $K_{\spatial\testing}$ and $K_{\testing\testing}$.
        We also introduce
        \begin{align*}
        \meanO_\idxi = \frac{1}{\idxL_\idxi}\sum_{\idxl=1}^{\idxL_\idxi}\sampleO_\idxi^\idxl,\qquad & \meanOall = (\overline{\sampleO}_1,\dots,\overline{\sampleO}_\idxI)^\top.
        \end{align*}
        Finally,
        \begin{align*}
        [ D ]_{\idxi\idxj}= \delta_{\idxi\idxj}&\frac{\varianceO}{\idxL_\idxi}\in\mathbb{R}^{\idxI\times\idxI},
        \end{align*}
        where $\delta_{\idxi\idxj}$ is the Kronecker delta (i.e.,
        $\delta_{\idxi\idxj}=1$ for $i=j$ and $\delta_{\idxi\idxj}=0$ otherwise). 
        \subsection{Distributed ML Estimator}
        The ML hyperparameter estimator is given by
        \begin{align}\label{eq:maximum-likelihood}
        \hyper_{\textnormal{ML}} = \argmax_{\hyper\in\hyperS} \prob(\sampleOall;\hyper).
        \end{align}
        
        We are now ready to state the first result characterizing the structure of
        optimization problem \eqref{eq:maximum-likelihood}.
        \begin{theorem}\label{thm:maximum-likelihood}
            The ML hyperparameter estimator $\hyper_{\textnormal{ML}}$ in
            \eqref{eq:maximum-likelihood} can be equivalently obtained by solving the
            following minimization problem
            \begin{equation}
            \min_{\hyper\in\hyperS} \;(\meanPallHyper -
            \meanOall)^\top(K_{\spatial\spatial} + D)^{-1}(\meanPallHyper - \meanOall).
            \label{eq:maximum-likelihood-centralized}
            \end{equation}
            \oprocend
        \end{theorem}
        The proof will be provided in a forthcoming document.

        Next, we introduce the distributed formulation of the optimization
        problem \eqref{eq:maximum-likelihood}.
        We start by observing that the most important consequence of \eqref{eq:sparsity}
        is that matrix $K_{\spatial\spatial}$ has the same sparsity as graph
        $\graph^\inter$ and so does $K_{\spatial\spatial} + \diag$. However, looking at
        equation \eqref{eq:maximum-likelihood-centralized}, we notice that
        $(K_{\spatial\spatial} + \diag)^{-1}$ is not necessarily a sparse matrix.
        In order to preserve the sparsity for a distributed computation, we then introduce
        a new optimization variable $\z$, so that $\hyper_{\textnormal{ML}}$ can
        be computed by solving the following optimization problem:
        \begin{alignat}{2}\label{eq:maximum-a-posteriori-z}
        \begin{split}
        \min_{\hyper\in\hyperS,\z\in\mathbb{R}^\idxI} \quad& \z^\top(K_{\spatial\spatial} + \diag)\z,\\
        \subj \quad& (K_{\spatial\spatial} + \diag)\z = \meanPallHyper - \meanOall.
        \end{split}
        \end{alignat}
        
        At this point, making explicit the relationship between $\meanPall$ and $\hyper$
        due to \eqref{eq:joint-assumption4}, we can write
        \eqref{eq:maximum-a-posteriori-z} as
        \begin{alignat}{2}\label{eq:maximum-a-posteriori-z-F}
        \begin{split}
        \min_{\hyper\in\hyperS,\z,\meanPall\in\mathbb{R}^\idxI} \quad& \z^\top(K_{\spatial\spatial} + \diag)\z,\\
        \subj \quad& (K_{\spatial\spatial} + \diag)\z = \meanPall - \meanOall,\\
        \quad& F_\idxi(\spatialBold_{\neigh_\idxi^\inter},\meanPall_{\neigh_\idxi^\inter};\hyper) = 0,\quad\idxi=1,\dots,\idxI.
        \end{split}
        \end{alignat}

        Defining the functions
        \begin{alignat*}{2}
        f_\idxi(\z_{\neigh_\idxi^\inter}) &=
        \frac{\varianceO}{\idxL_\idxi}\zScalar_\idxi^2 +
        \sum_{\idxj\in\neigh_\idxi^\inter} \zScalar_\idxi
        \covarianceP(\spatial_\idxi,\spatial_\idxj) \zScalar_\idxj,\quad&\idxi=1,\dots,\idxI,\\
        g_\idxi(\z_{\neigh_\idxi^\inter}) &=
        \frac{\varianceO}{\idxL_\idxi}\zScalar_\idxi + \overline{x}_\idxi +
        \sum_{\idxj\in\neigh_\idxi^\inter}\covarianceP(\spatial_\idxi,\spatial_\idxj)\zScalar_\idxj,\quad&\idxi=1,\dots,\idxI.
        \end{alignat*}
        we can rewrite \eqref{eq:maximum-a-posteriori-z-F} as
        \begin{alignat}{2}\label{eq:distributed-ml}
        \begin{split}
        \min_{\hyper\in\hyperS,\z,\meanPall\in\mathbb{R}^\idxI} \;\,& \sum_{\idxi=1}^\idxI f_\idxi(\z_{\neigh_\idxi^\inter}),\\
        \subj \;\,& g_\idxi(\z_{\neigh_\idxi^\inter}) = \meanP_\idxi,\\
        \;\,& F_\idxi(\spatialBold_{\neigh_\idxi^\inter},\meanPall_{\neigh_\idxi^\inter};\hyper) = 0,\;\idxi=1,\dots,\idxI.
        \end{split}
        \end{alignat}
        which is a formulation of problem \eqref{eq:maximum-likelihood-centralized}
        amenable to distributed solution.

        Due to this formulation, the solution
        $(\hyper_{\textnormal{ML}},\z_{\hyper_{\textnormal{ML}}},\meanPall_{\hyper_{\textnormal{ML}}})$
        of \eqref{eq:maximum-a-posteriori-z-F} can be computed by solving an
        optimization problem that has a separable cost (i.e., the sum of $\idxI$ local
        costs). Available distributed optimization algorithms 
        can be adopted to this aim, e.g., \cite{carli2015analysis},
        \cite{nedic2015distributed}, \cite{sun2016distributed}. The special sparsity of cost
        function and constraints could be exploited by using techniques as the one
        proposed in \cite{carli2013distributed}.
        
        \begin{remark}
            A special case is the one in which we have
            \begin{align*}
            F_\idxi(\spatialBold_{\neigh_\idxi^\inter},\meanPall_{\neigh_\idxi^\inter};\hyper) = F_\idxi(\spatialBold_{\neigh_\idxi^\inter},\meanP_{\idxi};\hyper),\quad\idxi=1,\dots,\idxI.
            \end{align*}
            In this case, if we define the functions
            \begin{align*}
            G_\idxi(\z_{\neigh_\idxi^\inter};\hyper) = F_\idxi(\spatialBold_{\neigh_\idxi^\inter},g_\idxi(\z_{\neigh_\idxi^\inter});\hyper),\qquad\idxi=1,\dots,\idxI,
            \end{align*}
            $\hyper_\textnormal{ML}$ can be computed by solving the optimization
            problem:
            \begin{alignat}{2}\label{eq:explicit-mean}
            \begin{split}
            \min_{\hyper\in\hyperS,\z\in\mathbb{R}^\idxI} \quad& \sum_{\idxi=1}^\idxI f_\idxi(\z_{\neigh_\idxi^\inter}),\\
            \subj \quad& G_\idxi(\z_{\neigh_\idxi^\inter};\hyper) = 0,\quad\idxi=1,\dots,\idxI,
            \end{split}
            \end{alignat}
            which has a simpler structure that can be exploited to speed up the calculation.
            Moreover, when $\meanP(\cdot)$ is an explicit function of $\hyper$, i.e.,
            $\meanP(\cdot) = \meanP(\cdot;\hyper)$, the constraints in
            \eqref{eq:explicit-mean} simplify as
            $G_\idxi(\z_{\neigh_\idxi^\inter};\hyper) = g_\idxi(\z_{\neigh_\idxi^\inter}) -
            \meanP(\spatial_\idxi;\hyper)$. \oprocend
        \end{remark}
        
        The vector $\meanPall_{\hyper_{\textnormal{ML}}}$ gives the evaluation of the
        Gaussian Process mean $\meanP(\cdot)$ in the measurement points. Once a value of
        $\hyper_{\textnormal{ML}}$ is available, computing nodes located at the
        measurement points may interpolate $\meanP(\cdot)$ at regression points
        $\testing_\idxm$ in their spatial proximity, i.e.,
        $\big[\meanPtestingAll_{\hyper_{\textnormal{ML}}}]_\idxm$, as follows.
        \begin{itemize}
            \item If $\meanP(\cdot)$ satisfies a (discretized) partial differential equation
            parametrized by $\hyper$, then
            $\big[\meanPtestingAll_{\hyper_{\textnormal{ML}}}]_\idxm$
            can be obtained by integrating the differential equation with hyperparameter
            $\hyper_{\textnormal{ML}}$ and suitable boundary conditions based on
            neighboring points in the interaction graph.
            \item If $\meanP(\cdot)$ is an explicit function of $\hyper$, then $\big[\meanPtestingAll_{\hyper_{\textnormal{ML}}}]_\idxm$ 
            can be obtained by
            $\big[\meanPtestingAll_{\hyper_{\textnormal{ML}}}]_\idxm =
            \meanP(\testing_\idxm;\hyper_{\textnormal{ML}})$.
            \item Alternatively, an interpolation can be performed by using the set of
            neighboring points.
        \end{itemize}

        \subsection{Local MAP Estimator}
        In this subsection we show how a node can locally compute the MAP estimator
        $\samplePtestingAll_{\textnormal{MAP}}$.
        Defining $\samplePtesting_\idxi = \sampleP(\testing_\idxi)$ and
        $\samplePtestingAll =
        (\samplePtesting_1,\dots,\samplePtesting_{\nTesting})^\top$,
        the MAP estimator in the regression points is given by
        \begin{align}\label{eq:maximum-a-posteriori}
        \samplePtestingAll_{\textnormal{MAP}} =
        \argmax_{\samplePtestingAll\in\spaceO^{\nTesting}\subset\mathbb{R}^M} \prob(\samplePtestingAll |
        \sampleOall; \hyper_{\textnormal{ML}}). 
        \end{align}
        Notice that, once a value for $\hyper_{\textnormal{ML}}$ is available, the
        estimation proceeds similarly as in a purely Bayesian set-up, where the prior
        distribution is \emph{fully specified} and given by
        $\mathcal{N}(\meanPtestingAll_{\hyper_{\textnormal{ML}}},
        K_{\testing\testing})$.
        Specifically, the case in which the prior is known and each sensor $\idxi$
        performs only one observation ($\idxL_\idxi = 1$) has been widely
        investigated in the literature \cite{rasmussen2006gaussian}.
        
        In our work we generalize this classical framework in two ways. First, we
        consider a heterogeneous network set-up in which nodes perform a different
        number of observations, so that even those with few observations take advantage
        of nodes with more observations (especially neighboring ones). Second, this
        MAP estimator exploits the ML estimation of the hyperparameters, thus
        ``optimally'' adapting the prior to all network data.
        In the next theorem we provide a closed form expression of the posterior
        distribution, and, based on that, we derive a formulation of the MAP estimator
        which can be computed by each node in the network in a decentralized way (i.e.,
        by collecting data from neighboring nodes in the interaction graph).
        
        \begin{theorem}\label{thm:maximum-a-posteriori}
            The posterior distribution
            $p(\samplePtestingAll|\sampleOall;\hyper_{\textnormal{ML}})$ is a multivariate
            Gaussian with mean
            $\meanPtestingAll_{\hyper_{\textnormal{ML}}} -
            K_{\testing\spatial}\z_{\hyper_\textnormal{ML}}$,
            where $\z_{\hyper_\textnormal{ML}}$ is a solution of problem
            \eqref{eq:distributed-ml}, and covariance matrix
            $K_{\testing\testing} - K_{\testing\spatial}(K_{\spatial\spatial} +
            \diag)^{-1} K_{\spatial\testing}$.
            Moreover, by defining
            $\mathcal{N}_\idxm^E := \{\idxj :
            \covarianceP(\testing_\idxm,\spatial_\idxj)\neq0\}$,
            the MAP estimator at $\testing_\idxm$ is given by
            \begin{align*} 
            \big[\samplePtestingAll_{\textnormal{MAP}}\big]_{\idxm} & =
            \big[\meanPtestingAll_{\hyper_{\textnormal{ML}}}]_\idxm
            - \sum_{\idxj\in\mathcal{N}_\idxm^E}
            \covarianceP(\testing_\idxm,\spatial_\idxj)\big[\z_{\hyper_\textnormal{ML}}\big]_\idxj. \eqoprocend
            \end{align*}
            
        \end{theorem}
        The proof will be provided in a forthcoming document.

        It is worth noting that if a regression point coincides with a
        measurement point, i.e., $\testing_\idxm = \spatial_\idxm$, then
        $\mathcal{N}_\idxm^E = \neigh_\idxi^\inter$ and
        \begin{align*}
        \big[\samplePtestingAll_{\textnormal{MAP}}\big]_{\idxi} =
        \big[\meanPall_{\hyper_\textnormal{ML}}\big]_\idxi -
        \sum_{\idxj\in\neigh_\idxi^\inter}\big[K_{\spatial\spatial}\big]_{\idxi\idxj}\big[\z_{\hyper_\textnormal{ML}}\big]_\idxj.
        \end{align*}
        For a regression point $\testing_\idxm$ that does not coincide with a
        measurement point, a computing node located at
        $\testing_\idxm$, or equivalently one
        of the network nodes in the proximity, can obtain
        $\big[\meanPtestingAll_{\hyper_{\textnormal{ML}}}]_\idxm$ through the
        interpolation procedure described in the previous subsection, and collect
        $\big[\z_{\hyper_\textnormal{ML}}\big]_\idxj$ from
        $\idxj\in\mathcal{N}_\idxm^E$. Notice that the sparsity of the kernel with
        respect to the entire space $P$ implies that $\mathcal{N}_\idxm^E$ typically
        contains a limited number of nodes.
        
        Generally speaking, Theorem \ref{thm:maximum-a-posteriori} provides an efficient way to compute the MAP estimator at each point. In fact, if one used the standard formula of the posterior (see \cite{rasmussen2006gaussian}) the MAP estimator would be $\meanPtestingAll_{\hyper_{\textnormal{ML}}} -
        K_{\testing q}(K_{qq} + \varianceO I_{L})^{-1}(\meanPall_L - \sampleOall)$ where $L = \sum_{i=1}^N L_i$, 
        $
        q = (\underbrace{s_1,\ldots,s_1}_{L_1},\ldots, \underbrace{s_N,\ldots,s_N}_{L_N})^\top
        $ and $\meanPall_L = (\underbrace{\mu(s_1),\ldots,\mu(s_1)}_{L_1},\ldots,\underbrace{\mu(s_N),\ldots,\mu(s_N)}_{L_N})^\top$, thus involving matrices of much larger size.

        \section{Numerical Simulations}
        \label{sec:simulations}
        In this section we analyze the numerical results related to the two example
        scenarios introduced in Section~\ref{subsec:examples}. We have used, in both
        cases, a kernel function known as Mahalanobis function (see
        \cite{melkumyan2009sparse}), which allows us to make $K_{\spatial\spatial}$
        sparse. Moreover, we took as regression points a fine grained discretization of
        the (one dimensional) set $\spatialS$ to better visualize the
        regression in the entire spatial domain.

        \subsection{Temperature Dynamics}
        
        For this numerical simulation, we have considered $\idxI = 10$ sensors uniformly
        spaced between $\spatial_1 = 0$ and $\spatial_\idxI = 2\pi/3$.
        The vector $\sampleOall$ of observations has been generated according to the following
        Temperature dynamics, 
        \begin{align*}
        \begin{cases} 
        \meanP''(\spatial) = -A\omega^2\sin(\omega\spatial + \phi), & \quad\spatial\in(\spatial_1,\spatial_\idxI), \\ 
        \hspace{2pt}\meanP(\spatial_1) = w_1, \\
        \meanP(\spatial_\idxI) = w_\idxI. 
        \end{cases}
        \end{align*}
        where the right hand side of the first equation represents the heat source with
        $A = 6$, $\omega = 3$ and $\phi = 3$, and the boundary conditions are $w_1 = 3$
        and $w_\idxI = 0$.
        
        This differential equation has a closed form solution
        $\theta(s) = A\sin(\omega\spatial + \phi) + C_1\spatial + C_0$, with  
        \begin{align*}
        C_1 &:= \frac{w_\idxI + A\sin(\omega\spatial_\idxI + \phi) - w_1 - A\sin(\omega\spatial_1 + \phi)}{\spatial_\idxI - \spatial_1},\\
        C_0 &:= w_1 - A\sin(\omega\spatial_1 + \phi) - C_1\spatial_1,
        \end{align*}
        so that the $\idxl$-th observation at $s_i$ is given by
        $x_\idxi^\idxl = \theta(s_\idxi) + \epsilon_\idxi$.
        
        In the estimation process we assume that the prior available information is the
        above family of Temperature dynamics with parameters
        $\hyper = (A,\omega,\phi)^\top$ being the hyperparameters of our framework. Once
        $\hyper_\textnormal{ML}$ has been obtained,
        $\meanPtestingAll_{\hyper_\textnormal{ML}}$ can be computed by integrating the
        Cauchy problem.
        
\begin{figure}
            \centering
            \includegraphics[scale=0.65]{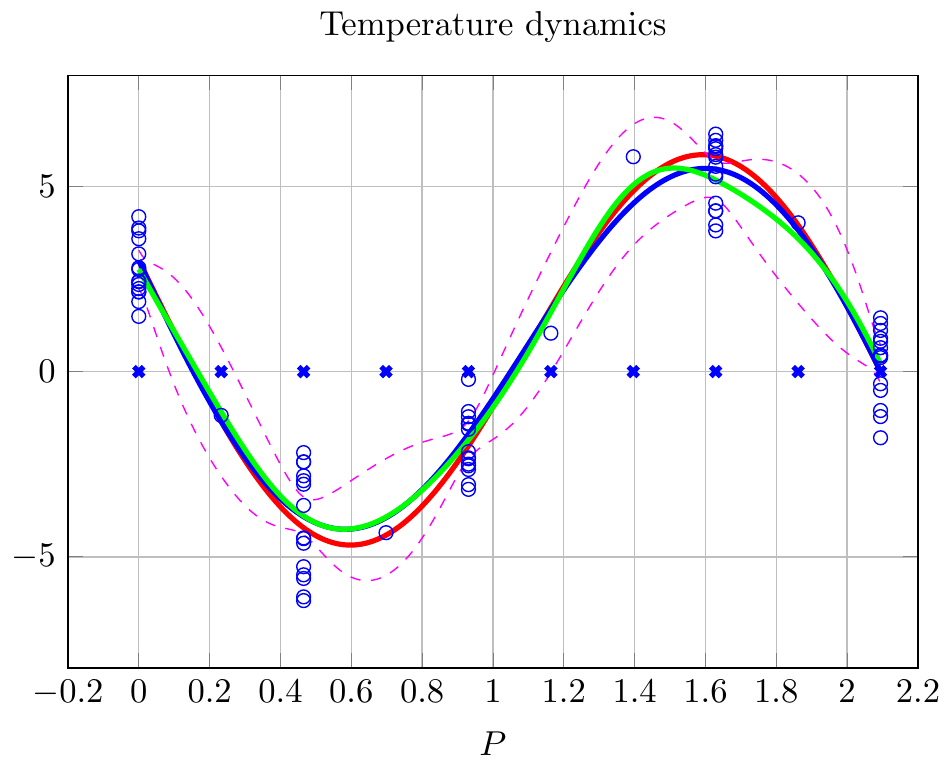}
            \caption{Temperature dynamics scenario: the measurement points
                $\spatial_\idxi$ are marked with a cross, while the observations
                $\sampleOall$ with circles. The red line is the true mean
                function $\meanP(\cdot)$, the blue line is
                $\meanPtestingAll_{\hyper_{\textnormal{ML}}}$ and the green line is
                the posterior mean $\samplePtestingAll_{\text{MAP}}$. In magenta we
                represent the 95\% confidence bound for the posterior
                mean.}\label{fig:poisson}
        \end{figure}

        The result of the simulation is depicted in Fig. \ref{fig:poisson}, where the
        estimated (process) prior mean in the regression points,
        $\meanPtestingAll_{\hyper_\textnormal{ML}}$, is depicted in blue, the posterior
        mean $\samplePtestingAll_{\text{MAP}}$ is shown in green and the true field
        $\theta(\cdot)$ in red.
        Remarkably, both the $\meanPtestingAll_{\hyper_\textnormal{ML}}$ and
        $\samplePtestingAll_{\text{MAP}}$ provide an accurate estimate of the spatial
        field $\sampleP(\cdot)$. As one would expect, at measurement points with a
        greater number of observations the local estimation is more accurate with a
        tighter confidence bound.
        
        \subsection{Interpolating Dynamics}
        In this case we have considered $\idxI = 12$ sensors spaced non-uniformly
        between $\spatial_1=-15$ and $\spatial_\idxI=14$. The vector $\sampleOall$ of
        observations has been generated according to a spatial field $\sampleP(\cdot)$
        being a quadratic polynomial, i.e.,
        \begin{align}\label{eq:true-mean}
        \sampleP(\spatial) = a\spatial^2 + b\spatial + c,
        \end{align}
        where we have chosen $a = 0.1$, $b = 0.1$ and $c = 10$.
        
        Differently from the previous numerical example, during the estimation process
        we have used as parametrized mean function (prior) an interpolating curve that
        does not match \eqref{eq:true-mean}. In fact we have used the family of
        \emph{natural cubic splines}. Specifically, we have taken $\idxI=12$ measurement
        points $s_1,\ldots,s_{12}$ clustered in five subregions. Then we have chosen as
        control points $c_j$ of the spline one point for each cluster, namely $c_1=s_1$,
        $c_2=s_3$, $c_3=s_5$, $c_4=s_7$ and $c_5=s_{12}$.
        
        Specifically, the spline function is defined as
        \begin{align*}
        \meanP(\spatial) = 
        \begin{cases}
        p_1(\spatial), & \mbox{ if } \spatial<c_1,\\
        p_j(\spatial), & \mbox{ if } \spatial\in[c_j,c_{j+1}),\\
        p_{\idxI-1}(\spatial), & \mbox{ if } \spatial\geq c_5,
        \end{cases}
        \end{align*}
        where each cubic polynomial $p_j(\cdot)$ satisfies the following conditions:  
        \begin{align*}
        \begin{cases}
        p_j(c_j) = \gamma_j,\quad p_j(c_{j+1}) = \gamma_{j+1}, & j=1,\dots,4,\\
        p_{j-1}'(c_j) = p_j'(c_j), & j=2,\dots,4,\\
        p_{j-1}''(c_j) = p_{j}''(c_j), & j=2,\dots,4\\
        p_1''(c_1) = 0,\quad p_{4}''(c_5) = 0,
        \end{cases}
        \end{align*}
        with $\hyper = (\gamma_1,\ldots,\gamma_5)^T$.
        
\begin{figure}
            \centering
            \includegraphics[scale=0.65]{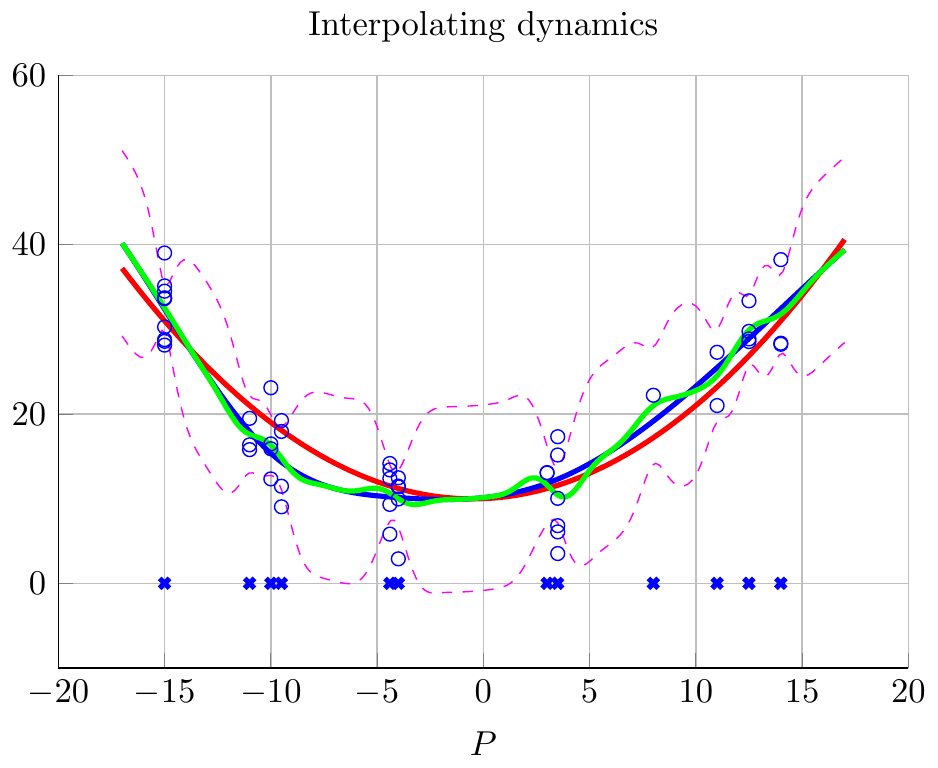}
            \caption{Interpolating dynamics scenario: the measurament points $\spatial_\idxi$ are labeled with a cross, while the observations $\sampleOall$ are marked with circles. The red line is the true mean function $\meanP(\cdot)$, the blue line is $\meanPtestingAll_{\hyper_{\textnormal{ML}}}$ and the green line is the posterior mean $\samplePtestingAll_{\text{MAP}}$. In magenta we represent the 95\% confidence interval relative to the posterior mean.}\label{fig:spline}
        \end{figure}

        The result of the simulation is depicted in Fig. \ref{fig:spline}. As in the
        previous example, the estimate is more accurate in those points with an higher
        number of observations. Moreover, we can appreciate the capability of the
        Interpolating dynamics of fitting the spatial field $\sampleP(\cdot)$ near the
        measurement points. 
        
        \section{Conclusions}
        
        In this paper we have proposed a novel probabilistic framework for distributed
        estimation of a spatial field in large-scale sensor networks, wherein each node
        performs multiple, local measurements.
        The proposed set-up is based on an Empirical Bayes approach in which the spatial
        field is modeled as a Gaussian Process with known covariance and unknown, but
        parametrized, mean. Specifically, we suppose that the mean function at the
        measurement points satisfies a set of equations (encoding some prior knowledge
        of the mean shape) parametrized by the unknown hyperparameters. 
        The Empirical Bayes estimation procedure consists of two parts: the computation
        of the ML estimate of the hyperparameters, and the computation of the MAP
        estimate of the field into regression points. In particular, we have shown
        that multiple observations improve the estimation accuracy by reducing the
        posterior variance, and we have also provided a sparse formulation of the ML
        optimization problem, which is amenable to distributed solution. Thus, each
        sensor optimally fuses information from the network in a distributed way thus
        improving its own MAP estimate of the field. 
        Numerical simulations for the proposed scheme have been presented for two
        example scenarios. In the first one, the temperature field in a thermostatic bar
        is estimated by using the Heat Equation as a prior knowledge for the mean. In the
        second one, interpolating curves have been used in a data-driven inference
        scenario with no physical knowledge of the field.

        \bibliographystyle{IEEEtran}
        \bibliography{distributed_kriging}

    \end{document}